\documentclass[aps,prl,showpacs,showkeys,twocolumn]{revtex4}

\newcommand{\beq}{\begin{equation}}
\newcommand{\eeq}{\end{equation}}
\newcommand{\bef}{\begin{figure}}
\newcommand{\eef}{\end{figure}}

\usepackage{graphicx}   
\usepackage{graphics}
\usepackage{amssymb}
\usepackage{pifont}

\begin{document}

\title{Symmetrical charge-charge interactions in ionic solutions: \\ implications for biological interactions}
\author{Eshel Faraggi}
\email{efaraggi@gmail.com}
\affiliation{Research and Information Systems, 155 Audubon Dr., Carmel, IN 46032, USA}
\date[]{Submitted to the Phys. Rev. Lett. 2009, PNAS-USA 2010, J. of Appl. Phys. 2011}

\begin{abstract}
As is well known in electrolyte theory, electrostatic fields are attenuated by the presence of mobile charges in the solution. This seems to limit the possibility of an electrostatic repulsion model of biological interactions such as cell division. Here, a system of two charges in an ionic solution is considered. It is found that in the context of the symmetries of the system, the electrostatic repulsion between the two is considerably increased as compared to the electrostatic repulsion between two bare charges in a dielectric. This increase in repulsion, directly resulting from interactions between the symmetrical parts of the system, was found to be  dependent on the magnitude of the charges and the separation between them. It was also found that this increases reaches a steady state for separation greater than a solvent determined length scale related to the Debye length. These findings strongly suggest that electrostatic interactions can play a crucial part in the physical forces that are involved in biological interactions.
 
\end{abstract}


\maketitle

\section{Introduction}
Biological cell division is the corner-stone of life. From a practical stand point, its malfunction is responsible for many diseases. Cancer and neural regeneration are two fields of research relevant to countless people that can gain tremendous advancements from a basic understanding of the physical mechanism of cell division and an ability to manipulate them. Even viral and bacterial infections require the division of cells and the propagation of genetic information. Because of cell division's primordial necessity for life, the fundamental underlying processes of it must be general. Because of these fundamental values, it is a candidate for systems where general biological interactions may be observed.

All life shares DNA as the basic code of its particular form. All life also shares the twenty amino-acids that make up proteins, the matter of life. This points in the direction of a common ancestor. It is also plausible to assume that the mechanisms for DNA segregation will rest on the same fundamental physics across different life forms. Probably the first thing the first living cells did was divide. Hence, it is required that the physical processes were simple, and did not involve complex biological machinery that would only develop later. If life originated by the chemicals of some ancient sea, acting freely under the forces of nature, those primordial cells were some form of dividing DNA. An extraordinary process.

There are four known fundamental forces. Two of them, the strong and weak nuclear forces, have to do with interactions on the atom's nucleus length scale. Per current knowledge, these forces do not contribute to life beyond their role in the creation of matter. A third force, gravity, has to do with large length scales, it is negligible in the biochemistry of living organism. This leaves us with the electromagnetic force as the sole interaction responsible for the differences between living and non-living matter. In fact, quantum  electromagnetism is the basis of atomic structure and interaction. When moving from a few atoms to the many atom systems prevalent in living organisms it is found that quantum effects become mostly negligible, and classical electromagnetism (EM) is used. Many models use EM to describe biological processes. In the realm of cell division, some form of EM must hide in the complexity.  

While it is known that tubulin, a protein related to cell division, posses a permanent dipole moment~\cite{mers04,scho05,rama06}, and that electric charges play a role in cell division~\cite{jaff77,mitbook,mitosis,hepl01,gagl02,fara06}, other types of models were also  proposed.~\cite{tayl63,nick88,glot05} A major objection to electrostatic models of cell division is Debye screening for ionic solutions.~\cite{tayl63,jackson} Considering the fields of a single charge in a solution of freely moving charges, one finds that the electrostatic potential is attenuated by an exponentially decaying envelope due to the presence of the mobile charges. The assumption being that the mobile charges will be distributed according to the Boltzmann distribution. Since the order of magnitude for the Debye screening length for biological matter is 1~nm, it was not expected that electrostatic repulsion can play a role in cell division. 

Here it will be shown that analysis of a symmetrical system, as is the case for example for the two kinetochores attached to the genetic code, shows an opposite trend. While the individual electromagnetic interaction is exponentially attenuated as prescribed by Debye screening, the many-body force of repulsion between the two symmetrical parts is increased as compared to bare charges. It is the intent of this work to argue that EM models can be constructed to describe cell division and other biological interaction. As briefly discussed above, there is no other candidate for a fundamental force except EM.

\section{Model}
Consider the Poisson equation for the potential $\Phi$, in a system of a point charge, $q_1$, at the origin, with stationary charged background (charge $-q_m$) and a cloud of mobile charged particles (charge $q_m$)
\beq
\nabla^2 \Phi = - 4 \pi q_1 \delta ({\mathbf x}) - 4 \pi q_m n_0 [ e^{-q_m \Phi / k_B T} - 1 ].
\label{debyeq}
\eeq
${\mathbf x}$ is the vector position, $n_0$ the particle density, and $k_B T$ the thermal energy. For a background of this equation see Jackson.~\cite{jackson} The second term on the right hand side of Eq.~(\ref{debyeq}) comes from the Boltzmann distributed mobile charges and the stationary ones. 

It is instructive to first examine charge conservation and the gauge invariance of the electrostatic potential. We need to separate two cases: a finite or infinite system. For a finite system a partition function is needed to normalize the exponential term in the Boltzmann distribution. This partition function will guarantee both the gauge invariance and conservation of charge. In the limit of infinite distances, the perturbation of the background/mobile charges due to the point charge should become negligible and hence
$$ e^{-q_m \Phi_{{\mathbf x} \rightarrow \infty } \, / k_B T} = 1.$$
This leads to the standard boundary conditions $\Phi_{{\mathbf x} \rightarrow \infty } =0$. A partition function normalization is possible if one wishes to add a constant to $\Phi$.

If one assumes that the EM energy is small compared to the thermal energy, one can approximate the term in the square brackets in Eq.~(\ref{debyeq}) as linear in $\Phi$ and the resulting differential equation leads to an exponentially decaying Coulomb field with a length scale of $1/e$ decay given by the Debye length. Since the Debye length is of the order of a few protein helix turns it was judged that electrostatic repulsion is of negligible importance in biological cell division.

\begin{figure}[th!]
\includegraphics[scale=0.5]{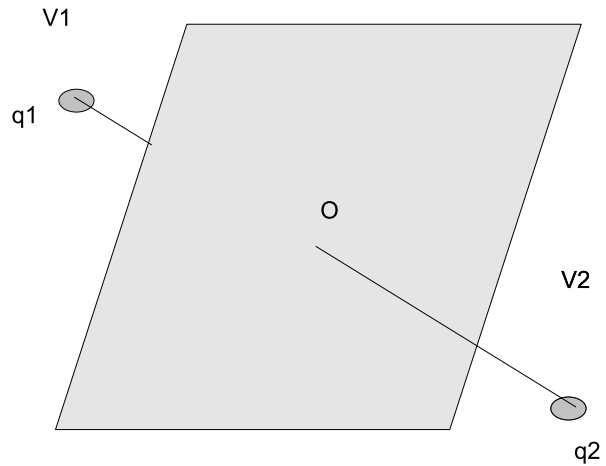}
\caption{Schematic representation of the model. Two charges, $q_1$ and $q_2$, are separated by the colored plane in the figure which is perpendicular to the line joining the charges and placed equidistant between them. This plane defines two volumes, $v_1$ and $v_2$ as shown in the figure.}
\label{fig_2vols}
\end{figure}

The crucial point that this work picks up on is that the dividing biological cell is inherently a symmetric process. This implies that we are interested in the interaction between two charges in such an environment. As will be discussed in a future, more biologically oriented publication, the existence of the two charges can come about from the dipolar nature of tubulin. Here the focus is on the physical interaction of a system as depicted in Fig.~\ref{fig_2vols}. A mirror symmetry is taken between the two parts of the system. 

The modified Poisson equation for the two charges, $q_1 = q_2 = q$ immersed in an aqueous solution with dielectric $D$, is given by
\beq
D \cdot \nabla^2 \Phi = - 4 \pi q [ \delta ({\mathbf x}+{\mathbf x}_0) + \delta ({\mathbf x}-{\mathbf x}_0)] - 4 \pi q_m n_0 [ e^{-q_m \Phi / k_B T} - 1 ],
\label{debyeq2}
\eeq
taking a coordinate system where the charges are at $\pm {\mathbf x}_0$ as described in Fig.~\ref{fig_2vols}. As before, if $|q_m \Phi / D k_B T|<<1$ one can approximately solve Eq.~(\ref{debyeq2}) and obtain
\beq
\Phi ({\mathbf x}) = \frac{q}{D} [ \frac{e^{- k_D || {\mathbf x}-{\mathbf x}_0 ||}}{|| {\mathbf x}-{\mathbf x}_0 ||} + \frac{e^{- k_D || {\mathbf x}+{\mathbf x}_0 ||}}{|| {\mathbf x}+{\mathbf x}_0 ||}],
\label{debyeq2_sol}
\eeq
with $k_D = 4 \pi n_0 q_m^2 \, / \, D k_B T$ the Debye constant in a dielectric.

This is the field between the two charges, attenuated by Debye screening. For a homogeneous system the electromagnetic fields far away from the two charges is attenuated. Here the interest is in the force of separation between the charge densities of the two volumes delineated by the symmetry of this problem.  As displayed in Fig.~\ref{fig_2vols} these volumes are separated by the plane containing the origin and perpendicular to the line joining $q_1$ and $q_2$. The Coulomb force between the volumes $V_1$ and $V_2$ can be formally written as
\beq
F_s = \int_{{\mathbf x} \in V_1} \int_{{\mathbf x}' \in V_2} \frac{\rho({\mathbf x}) \rho({\mathbf x}')}{D \, || {\mathbf x} - {\mathbf x}' ||^2} d^3{\mathbf x}' \, d^3{\mathbf x},
\label{fofsep}
\eeq
with
$$
\rho({\mathbf x}) = q [ \delta ({\mathbf x}+{\mathbf x}_0) + \delta ({\mathbf x}-{\mathbf x}_0)] + q_m n_0 [ e^{-q_m \Phi / k_B T} - 1 ].
$$

The integral in Eq.~(\ref{fofsep}) can be re-represented to aid in its interpretation. Integrating out the delta functions, defining $\gamma ({\mathbf x}) = e^{-q_m \Phi ({\mathbf x}) / k_B T} - 1$, and using the symmetry of the system  $F_s$ becomes
\beq
F_s = F_b + F_a + F_r
\label{fofseq_red} 
\eeq
with
\beq
F_b = \frac{q^2}{D \, || 2 {\mathbf x}_0 ||^2}
\eeq
\beq
F_a = \frac{2 n_0 q_m q}{D} \int_{{\mathbf x} \in V_1} \frac{\gamma ({\mathbf x})}{|| {\mathbf x} - {\mathbf x}_0 ||^2} \, d^3{\mathbf x}
\eeq
\beq
F_r = \frac{(n_0 q_m)^2}{D} \int_{{\mathbf x} \in V_1} \int_{{\mathbf x}' \in V_2} \frac{\gamma ({\mathbf x}) \gamma ({\mathbf x}')}{|| {\mathbf x} - {\mathbf x}' ||^2} \, d^3{\mathbf x} \, d^3{\mathbf x}'
\eeq

Equation~(\ref{fofseq_red}) breaks up the interaction forces into three components. $F_b$ is the Coulomb force between the two charges external to the ionic solution. $F_a$ represents the force between the induced ionic charges and the bare charge in the opposite volume. Specifically, it represents the interaction of charge 1 with the ionic cloud surrounding charge 2 plus the interaction of charge 2 with the ionic cloud surrounding charge 1. Finally, $F_r$ represents the force between the induced ionic charges in the two volumes. Since EM is a linear theory, the total force between the two volumes is the superposition of the individual forces. The symmetry of the system considered dictates that both $F_b$ and $F_r$ are repulsive, while $F_a$ is attractive.
  
Up to now the derivation for the two particle system is general in terms of the relationship between thermal and EM energies. In principle, if one can solve Eq.~(\ref{debyeq2}) and obtain $\gamma$, one can use Eq.~(\ref{fofseq_red}) to obtain the force between the charge distribution in the daughter cells. One should note that the general solution is not symmetric with respect to charge. This comes about since for an infinite system we are unlimited in the amount of mobile charges we can add. Positive charge density, on the other hand, is introduced by the removal of this mobile charges. Hence, it is bounded by the initial, unchanging density of the immobile ones.

In practice, however, it is well known that biologically relevant ionic energies are significantly smaller than thermal ones. One major reason for this is the abundance of water molecules. Due to their polarity these tend to aggregate around charges and limit the access distance for ions.
If one assumes that the ionic energy is small compared to the thermal energy, one can approximate $ \gamma $ as $ -q_m \Phi / k_B T $. For convenience, define from Eq.~(\ref{debyeq2_sol})  $\psi =D \,  \Phi / q$, this enables us to write for $F_a$ and $F_r$:
\beq
F_a = - 2 \frac{n_0 (q_m q)^2}{D^2 k_B T} \int_{{\mathbf x} \in V_1} \frac{\psi ({\mathbf x})}{|| {\mathbf x} - {\mathbf x}_0 ||^2} \, d^3{\mathbf x},
\label{fa_aprx}
\eeq
\beq
F_r = \frac{(n_0 q_m^2 q)^2}{D^3 (k_B T)^2} \int_{{\mathbf x} \in V_1} \int_{{\mathbf x}' \in V_2} \frac{\psi ({\mathbf x}) \psi ({\mathbf x}')}{|| {\mathbf x} - {\mathbf x}' ||^2} \, d^3{\mathbf x} \, d^3{\mathbf x}'.
\label{fb_aprx}
\eeq

In this regime, changes in the charge density are small, and as exhibited by Eqs.~(\ref{fa_aprx},\ref{fb_aprx}) are symmetric with respect to charge reversal. If the charge density fluctuations are larger this symmetry is lost, as mentioned previously. We should now consider the finite size screening of the EM forces near the bare charges. Since this involves the collective behavior of a large number of water  molecules it is an entropic effect and will depend on the temperature. For convenience it will be assumed that $q_m \Phi \le 0.25 * k_B T$, with $T = 300$K. This condition is also necessary to satisfy the requirements of the linear approximation for $\gamma$. A physical interpretation for this condition is a layer of water of radius $a = q q_m / 2 \cdot 0.25 D k_B T$ surrounding the charges. This layer inhibits closer approach between the various charges. Thus, to the edge of the water layer, EM fields propagate through an ion-less dielectric. That is, Eq.~(\ref{debyeq2_sol}) should be modified to separate distances smaller and larger than $a$. Specifically, the EM field decays exponentially only for distances greater than $a$ and a continuous field is assumed at $a$. This amounts to subtracting $a$ from the exponent in Eq.~(\ref{debyeq2_sol}) for the potential outside the dielectric region.

\section{Results}
\begin{figure}[th!]
\includegraphics[scale=0.5]{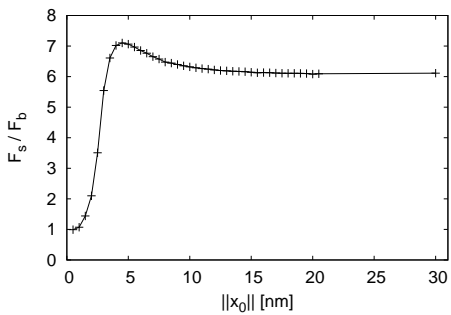}
\caption{Ratio between volume separation force, $F_s$, and the force between two proton charges in a dielectric, $F_b$, as a function of the separation between the charges. We see that for $||{\mathbf x}_0|| < a = 2.7$~nm the ratio increases as a the separation increases and it  reaches saturation of around 6.5 for $||{\mathbf x}_0|| \approx 4$~nm, corresponding with $a$. The curve through the data points is used merely to aid the eye.}
\label{res1}
\end{figure}

The analysis of the treatment outlined above was done using trapezoidal  numerical integration coded in FORTRAN. Given on the y-axis of  Fig.~\ref{res1} is the total force between the charge distributions in the two daughter cells, $F_s$, normalized by the Coulomb force between two core charges taken as protons in a water dielectric, $F_b$. On the x-axis of Fig.~\ref{res1} the distance between the charges is given. We see that up to a separation of approximately 3~nm $F_s \approx F_b$. This is due to the limit we put on the distance between the charges which for this case is given by 2.7 nm. As the separation is increased, effects from the ionic field become more noticeable with ions in the two daughter cells repealing each other. This regime of sharp transition, where $F_s$ becomes almost an order of magnitude larger than $F_b$, reaches a plateau around 4 nm after which $F_s \rightarrow 6 \cdot F_b$. That is, due to the symmetry between the charge distribution in the two daughter cells the repulsive force between $q_1$ and $q_2$ is increased six-fold as compared to their Coulomb repulsion and this ratio remains approximately constant over the range of the calculation. 

\begin{figure}[th!]
\includegraphics[scale=0.5]{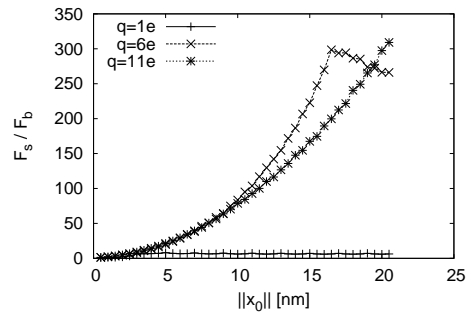}
\caption{Ratio between  $F_s$ and $F_b$, as a function of the separation between the charges. In this case three q-values are presented. The previous case of $q=1e$ is barely discernible on this scale. The case of $q=6e$ reaches saturation ratio of about 300. The case of $q=11e$ does not reach saturation since for this case $a=30$~nm.}
\label{res2}
\end{figure}

Results for $q=1,6,11 \, e$ with $e$ the charge of the proton are given in Fig.~\ref{res2}. Again, we see that for small separation $F_s \approx F_b$. As the separation is increased, contribution to the repulsive force from the ionic field become more notable. These are most likely due to contributions from the cleavages in the misaligned spheres defined by $a$ around $q_1$ and $q_2$. For $q=11e$, $a = 30$~nm, explaining the inability of this case to saturate by $||{\mathbf x}_0|| =$20~nm, the largest calculated separation. For the case of $q=6\;e$, $a = 16$~nm, which corresponds well with the saturation point for that curve. Note that the boundary condition for the numerical integration were set at 31~nm due to computational reasons. The proximity of the boundaries causes the decay observed in Fig.~\ref{res2} for the $q=6e$ case past saturation. A similar effect was observed for $q=1e$. Note that the single point in Fig.~\ref{res1} at $||{\mathbf x}_0|| = 30$~nm was calculated with the outer boundary at 61~nm.  This boundary effect also critically limits the accuracy of the results for $q=11e$, and is exhibited by the crossing of the lines for $q=11e$ and the $q=6e$ at a separation of approximately 8~nm, due to the greater $a$ of the $q=11e$ case.

\section{Discussion}
A system of two charges in an ionic solution was considered. It was found that in the context of Debye ionic screening the two symmetrical parts delineated by these charges experience an effect that increases the electrostatic repulsion between them as compared to the ion-free case. This increase in repulsion is dependent on the magnitude of the charges and the separation between them. It was also found that this effect reaches a steady state for separations greater than a solvent determined length scale related to the Debye constant. These findings strongly suggest that electrostatic interactions can play a crucial role in the physical forces that are responsible for biological interactions such as cell division. It is also interesting that other deviations from electrostatic interactions were observed, for example for the system of lipid membranes.~\cite{petr06}

A separate question, under current investigation, is that of the mechanisms and processes that would be able to reproduce the observed symmetrical configuration of the dividing biological cell, and their relationship to the  configurations discusses here. The main ingredients of this formulation are: dipolar tubulin, symmetry in its connections to the centrosome, and the kinetochore. Since the tubulin connects with a preferential polarity to the centrosome, through the dipole-dipole interaction along the tubulin fiber, charges of identical polarity will connect to each of the kinetochores. These two repulsing charges in the kinetochores could be the basis of the separation of DNA and biological matter.

\begin{acknowledgments}
The author would like to gratefully acknowledge the continuous support of Natali Teszler.
\end{acknowledgments}

\bibliographystyle{unsrt}

\begin{thebibliography}{10}
\bibitem{fn1}
{Note that the term ``daughter cell'' is used rather loosely here and refers best to a primordial blob of splitting biological matter, possibly akin to binary fission.}
\end{thebibliography}

\end{document}